\documentclass[aps,prl,twocolumn,showpacs,superscriptaddress, longbibliography]{revtex4-1}
\usepackage{graphicx}  
\usepackage{dcolumn}   
\usepackage{bm}        
\usepackage{amssymb}   
\usepackage{natbib}
\usepackage{amsmath}
\usepackage{xcolor}
\begin{document}
\raggedbottom

\title{Programmable interference between two microwave quantum memories}
\author{Yvonne Y. Gao}
\email[corresponding author:]{yvonne.gao@yale.edu}
\affiliation{Departments of Physics and Applied Physics, Yale University, New Haven, Connecticut 06520, USA}
\affiliation{Yale Quantum Institute, Yale University, New Haven, Connecticut 06511, USA}
\author{B.J. Lester}
\affiliation{Departments of Physics and Applied Physics, Yale University, New Haven, Connecticut 06520, USA}
\affiliation{Yale Quantum Institute, Yale University, New Haven, Connecticut 06511, USA}
\author{Yaxing Zhang}
\affiliation{Departments of Physics and Applied Physics, Yale University, New Haven, Connecticut 06520, USA}
\affiliation{Yale Quantum Institute, Yale University, New Haven, Connecticut 06511, USA}
\author{C. Wang}
\affiliation{University of Massachusetts, Amherst, MA 01003-9337 USA}
\author{S. Rosenblum}
\affiliation{Departments of Physics and Applied Physics, Yale University, New Haven, Connecticut 06520, USA}
\affiliation{Yale Quantum Institute, Yale University, New Haven, Connecticut 06511, USA}
\author{\\L. Frunzio}
\affiliation{Departments of Physics and Applied Physics, Yale University, New Haven, Connecticut 06520, USA}
\affiliation{Yale Quantum Institute, Yale University, New Haven, Connecticut 06511, USA}
\author{Liang Jiang}
\affiliation{Departments of Physics and Applied Physics, Yale University, New Haven, Connecticut 06520, USA}
\affiliation{Yale Quantum Institute, Yale University, New Haven, Connecticut 06511, USA}
\author{S.M. Girvin}
\affiliation{Departments of Physics and Applied Physics, Yale University, New Haven, Connecticut 06520, USA}
\affiliation{Yale Quantum Institute, Yale University, New Haven, Connecticut 06511, USA}
\author{R.J. Schoelkopf}
\email[corresponding author:]{robert.schoelkopf@yale.edu}
\affiliation{Departments of Physics and Applied Physics, Yale University, New Haven, Connecticut 06520, USA}
\affiliation{Yale Quantum Institute, Yale University, New Haven, Connecticut 06511, USA}
\date{\today}

\begin{abstract}
Interference experiments provide a simple yet powerful tool to unravel fundamental features of quantum physics. Here we engineer an RF-driven, time-dependent bilinear coupling that can be tuned to implement a robust 50:50 beamsplitter between stationary states stored in two superconducting cavities in a three-dimensional architecture. With this, we realize high contrast Hong-Ou-Mandel (HOM) interference between two spectrally-detuned stationary modes. We demonstrate that this coupling provides an efficient method for measuring the quantum state overlap between arbitrary states of the two cavities. Finally, we showcase concatenated beamsplitters and differential phase shifters to implement cascaded Mach-Zehnder interferometers, which can control the signature of the two-photon interference on-demand. Our results pave the way toward implementation of scalable boson sampling, the application of linear optical quantum computing (LOQC) protocols in the microwave domain, and quantum algorithms between long-lived bosonic memories. 
\end{abstract}

\pacs{}
\maketitle
\section{I. Introduction}
Interference experiments are one of the simplest probes into many of the riveting facets of quantum mechanics, from wave-particle duality to non-classical correlations The seminal work by Hong, Ou, and Mandel (HOM) is an elegant manifestation of two-particle quantum interference arising from bosonic quantum statistics~\cite{hom1987}. In their experiment, two photons incident on a 50:50 beamsplitter always exit in pairs from the same, albeit random, output port. Central to such interference experiments is the unitary operation $\hat{U}_{\mathrm{BS}}=\mathrm{exp}[i\frac{\pi}{4}(\hat{a}\hat{b}^\dagger + \hat{a}^\dagger\hat{b})]$. For propagating particles, this is simply realized with a 50:50 beamsplitter, but more generally, it can be implemented by engineering a time-dependent coupling of the form $\hat{H}_{\mathrm{int}}(t)/\hbar=g(t)\hat{a}\hat{b}^\dagger + g^{*}(t)\hat{a}^\dagger\hat{b}$. Recent experiments have demonstrated this type of coupling in different physical systems, enabling interference of both bosonic and fermionic particles~\cite{kaufman2014, bocquillon2013, lopes2015,toyoda2015}. These results have shed light on the concept of entanglement~\cite{tichy2014, islam2015} and enabled fundamental tests of quantum mechanics like the violation of Bell's inequalities~\cite{aspect1999}. They also have profound technological implications,with new applications in areas such as quantum metrology~\cite{kimble2001}, simulation~\cite{leibfried2002}, and information processing~\cite{knill2001,kok2007}. 

\begin{figure}[!hbt]
\centering
\includegraphics[scale=1]{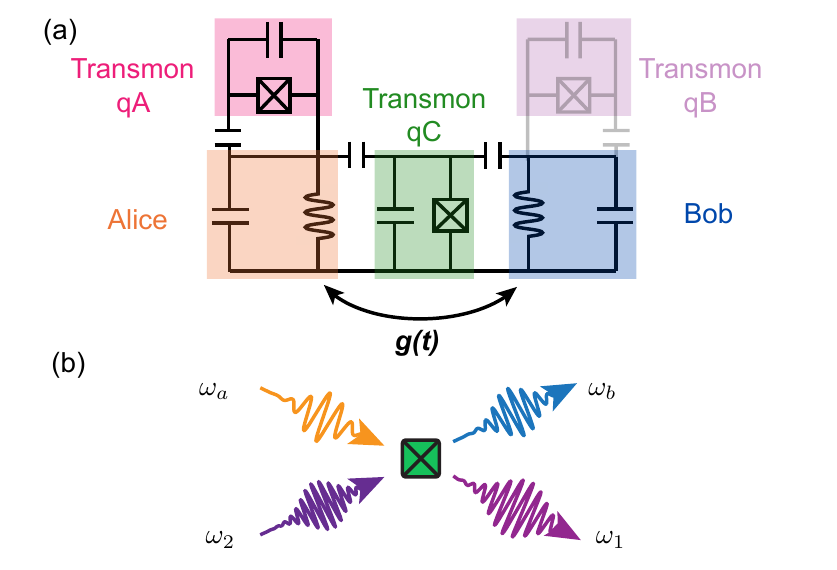}
\caption{\textbf{cQED system.} (a) The effective circuit of the cQED system containing two high-Q harmonic oscillators (orange, blue) bridged by  a transmon (green), as well as an additional transmon mode that only capacitively couples to Alice (pink). A similar ancillary transmon (purple) can also be used, but it is not necessary for this experiment. The resonance frequency of Alice and Bob are detuned from each other by $\sim$ 1 GHz to minimize undesired cross-talk. (b) Four-wave mixing process through the Josephson junction of qC that enables the bilinear coupling between Alice and Bob when $\omega_2 - \omega_1 = \omega_{b} - \omega_{a}$. }
\label{fig:circuit}
\end{figure}

Superconducting systems have been proposed as a promising platform to study bosonic interference and implement scalable boson sampling~\cite{Peropadre2016}. In particular, superconducting microwave cavities coupled to transmon or other non-linear ancillae in the circuit quantum electrodynamics (cQED) framework have the capability to deterministically create complex bosonic states~\cite{hofheinz2008, heeres2015} and perform robust measurements of photon statistics. So far, interference between harmonic oscillator modes has been demonstrated in cQED systems using elements with tunable frequencies~\cite{lang2013, Nguyen2012, Mariantoni2011, pierre2018}. However, such systems tend to suffer from unfavorable coherence properties, limiting the complexity of the experiment to single photons. Recently it has been shown that three-dimensional (3D), fixed-frequency superconducting microwave cavities have excellent coherence properties~\cite{reagor2013, romanenko2017}, making them attractive quantum memories. We can also readily prepare and manipulate complex quantum states in these cavities using the transmon as an ancilla. However, direct interference between bosonic states stored in these memories has remained a challenge due to the complexity of realizing high quality beamsplitters (BS)  and differential phase shifters (DPS) between stationary quantum memory modes. 

In this work, we showcase the on-demand interference of stationary bosonic modes stored in two spectrally-separated, long-lived superconducting microwave cavities, Alice and Bob. Our implementation employs a frequency-converting bilinear coupling between them which can be programmed to effectively implement a robust BS. With this capability, we demonstrate HOM interference between the two memories with a contrast up to $98\%\pm1\%$. Further,  we combine this with photon number parity measurement to perform efficient determination of quantum state overlap~\cite{filip2002, islam2015}. Lastly, we demonstrate in-situ manipulation of two photon interference through cascaded Mach-Zehnder (MZ) interferometers constructed with multiple BS and DPS. This highlights the versatility of our implementation and opens the door to more complex interference experiments in cQED. 

\section{II. Implementation and characterization}
Stationary photons stored in the two spectrally-separated cavities can only interfere if their energies are made indistinguishable. To do so, we can engineer a direct, tunable bilinear coupling $\hat{H}_{\mathrm{int}}$ between Alice and Bob, whose resonance frequencies are separated by $\sim1\,\mathrm{GHz}$ to ensure minimal residual coupling~\cite{supplement}. The effective circuit of the system is depicted in Fig.~\ref{fig:circuit}(a). The transmon ancilla, qA, couples only to Alice to provide the capability of fast single cavity manipulations. The Y-shaped transmon~\cite{wang2016}, qC, weakly couples dispersively to both Alice and Bob. Its single Josephson junction provides the necessary nonlinearity to activate the desired bilinear coupling through four-wave mixing in the presence of the appropriate drives. The Hamiltonian of the Josephson junction in presence of two drives, with normalized amplitudes $\xi_{1}$ and  $\xi_{2}$, is given by~\cite{nigg2012, leghtas2015}
\begin{align}\label{eq:cosine}
\hat{H} = - E_{J}\cos{[\phi_{a} (\hat{a} + \hat{a}^{\dagger}) +\phi_{b} (\hat{b} + \hat{b}^{\dagger})} \nonumber \\
+ \phi_{c} (\hat{c} + \hat{c}^{\dagger} + \xi_1 + \xi^{*}_1 + \xi_2 + \xi^{*}_2)]
\end{align}
where $\hat{a}$ and $\hat{b}$ are the creation operators of the two harmonic oscillator modes respectivley, $\hat{c}$ is that of qC, $E_{J}$ is the Josephson energy of qC, and $\phi_{i} (i=a, b, c)$ is the zero-point fluctuation of the phase associated with Alice, Bob, and qC respectively. 

\begin{figure}[!htb]
\centering
\includegraphics[scale=1]{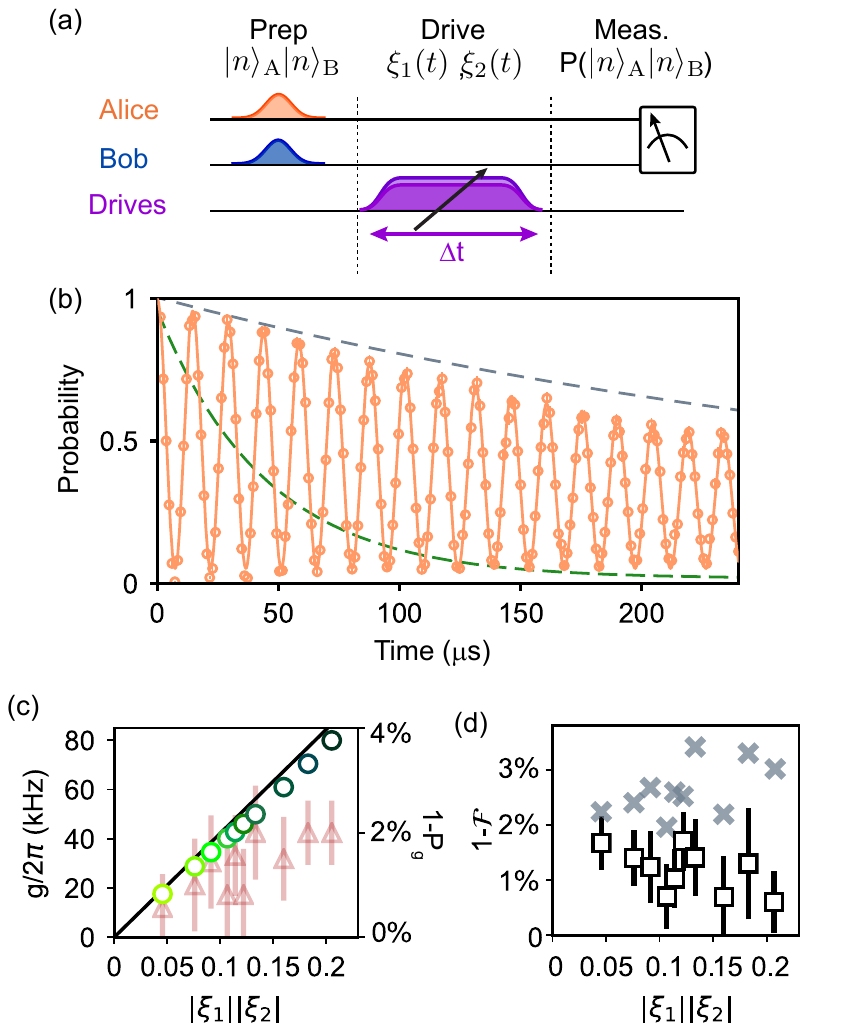}
\caption{\textbf{Calibration of bilinear coupling.} (a) General measurement protocol. After preparing in the cavities in the desired initial state, we apply the drives for a variable amount of time before measuring the final joint memory state using a photon number selective $\pi$ pulse on qC. (b) Measured $P_{10}$ (orange circles) with the engineered interaction at $|\xi_{1}||\xi_{2}| \approx 0.12$. Data are normalized by a constant scaling factor to calibrate out the SPAM errors~\cite{supplement}. Solid line shows fit to the functional form $P_{10} \propto e^{t/\tau_{1}}[1+ e^{t/\tau_{\phi}}\sin{(2\pi t f})]$. Dashed lines are the intrinsic decays of $|1\rangle_{\mathrm{A}}$ (grey) and relaxation of $|e\rangle$ (green) of qC. (c) Measured coupling strength $g/2\pi$ (green circles) as function of the drive strength $|\xi_{1}||\xi_{2}|$. Data is consistent with the predicted behaviour (black line) based on the 4th order cosine expansion, but deviate at higher drive powers. Measured excited state population of qC (red triangles), after a single BS operation. (d) Measured infidelity of a BS as defined by $\tau_{BS}/T_{BS}$ at different drive powers with (black squares) and without (grey crosses) post-selecting on qC remaining in its ground state after the operation.}
\label{fig:calibrations}
\end{figure}

The desired bilinear coupling is realized by supplying two RF drives whose detunings match that between Alice and Bob. The resulting interaction Hamiltonian can be written as~\cite{supplement}: 
\begin{equation}
\hat{H}_{\mathrm{int}}(t)/\hbar= g(t)(e^{i\varphi}\hat{a}\hat{b}^{\dagger} + e^{-i\varphi} \hat{a}^{\dagger}\hat{b})
\end{equation}
where $\varphi$ is determined by the relative phases of the two drives, and the coupling coefficient is $g(t)=E_{J}\phi_a\phi_b\phi^2_c\xi_{1}(t)\xi_{2}(t)=\sqrt{\chi_{ac}\chi_{bc}}\xi_{1}(t)\xi_{2}(t)$~\cite{supplement, zhang2018}. The strength of each drive $\xi_{1, 2}$ is calibrated independently by measuring the Stark shift of the resonance frequency of qC. The dispersive couplings $\chi_{ac}$ and $\chi_{bc}$ are determined using standard number-splitting measurements~\cite{supplement}. This coupling between two harmonic modes has been shown~\cite{pfaff2017} to transfer a quantum state from a memory to a propagating mode. Here, we engineer the same coupling but between two high-Q modes so that we can realize the unitary operation $U(\theta) = \mathrm{exp}[-\frac{i}{\hbar}\int_{0}^{T} H_{\mathrm{int}}(t)dt]$ while only virtually populating the excited levels of qC.  We define $\theta = \int_{0}^{T} g(t) dt$ as is the effective mixing angle~\cite{Nguyen2012} of the process. It is fully tunable by varying the duration of the RF drives. For $\theta = \frac{\pi}{2}$ (mod $\pi$), the unitary performs a SWAP operation that exchanges the states between the two memories, while for $\theta = \pi/4$ (mod $\pi/2$) it corresponds to a 50:50 BS operation ($\hat{U}_{\mathrm{BS}}$). 

We calibrate the strength of the engineered coupling by monitoring the dynamics of a single excitation under $\hat{U}(\theta)$. As shown in Fig.~\ref{fig:calibrations}(a), we initialize the memories in $|1, 0\rangle_\mathrm{AB}$ using numerically optimized pulses~\cite{heeres2017} while ensuring that qC remains in $|g\rangle$. We then apply the drives for a variable duration before measuring the joint population distribution in Alice and Bob using a photon-number selective $\pi$ pulse on qC~\cite{schuster2007}. Using this method, we monitor the evolution of the single excitation as it coherently oscillates between the two memories $\sim100$ times faster than their photon loss rates. An example is shown in Fig.~\ref{fig:calibrations}(b) at a coupling strength $g/2\pi = 34$ kHz. This corresponds to implementing a BS operation in $T_{\mathrm{BS}} = \pi/4g \approx$ 3.6\,$\mu$s, two orders of magnitude faster than the natural coupling ($\sim 1/\chi_{ab}$) between the two detuned memories~\cite{supplement}. Thus, this ensures that the operation has a large on-off ratio. 

We can assess the fidelity of the BS operation by analyzing the decoherence time associated with the evolution of a single excitation under $\hat{U}_{\theta}$. Two mechanisms could introduce non-idealities to the operation, namely, photon loss and dephasing. By summing the measured $P_{10}$ and $P_{01}$, we obtain an envelope whose exponential decay gives the effective relaxation time $\tau_{1}$.  We can then divide $P_{10}$ by this envelope to extract the dephasing time $\tau_{\phi}$ using a decaying sinusoidal fit. For the data shown in Fig.~\ref{fig:calibrations}b, we extract an effective $\tau_{1} \simeq 400\,\mu$s and $\tau_{\phi} \simeq 800\,\mu$s at $|\xi_{1}||\xi_{2}| \approx 0.12$. Both are consistent with independent measurements of the coherence times of Alice and Bob. Combining these, we infer an effective decoherence time $1/\tau_{\mathrm{BS}}= 1/2\tau_1 + 1/\tau_{\phi} \approx$ 400\,$\mu$s for the operation with an infidelity of $\sim 1\%$ obtained by comparing this to the time required to implement $\hat{U}_{\mathrm{BS}}$ at this drive power. 

We then characterize the coupling strength and the fidelity of the BS at different drive powers. We show in Fig.~\ref{fig:calibrations}(c) that $g$ scales linearly with $|\xi_{1}||\xi_{2}|$ at low powers, consistent with the predictions from the 4th order expansion of Eq.~\ref{eq:cosine}. However, it deviates from this simple model when the drives are stronger, which can be explained by a perturbation theory treatment~\cite{supplement}. Naively, one might think that the decoherence would contribute less to the infidelity as the operation speeds up. However, as $g$ increases so does the participation of the qC excited levels, which are measured independently after each BS. We observe that the probability of qC departing from its ground state $1-P_{g}$ increases from $0.6\%$ to $2\%$ as we increase the drive strengths. The effect of this is two fold: an apparent reduction in the readout contrast as well as faster decoherence during $\hat{U}_{\mathrm{BS}}$. The former can be mitigated by performing post-selection: data are discarded if qC does not remain in $|g\rangle$ after the operation. This ensures that qC is a faithful meter for the joint photon population of Alice and Bob. We attribute the degradation of coherence to the greater participation of qC. This causes the system to inherit less favorable coherence times during $\hat{U}_{\mathrm{BS}}$~\cite{supplement}. Combining the faster operations and the penalty due to increased qC participation, the overall infidelity ends up roughly constant over different drive powers as shown in Fig.~\ref{fig:calibrations}(d). For subsequent experiments, we operate at $|\xi_{1}||\xi_{2}| \simeq 0.1$ where the drives do not introduce measurable non-idealities on top of the intrinsic decoherence of the system. 

\section{III. Interference between two microwave quantum memories}
\begin{figure}[!hbt]
\centering
\includegraphics[scale=1]{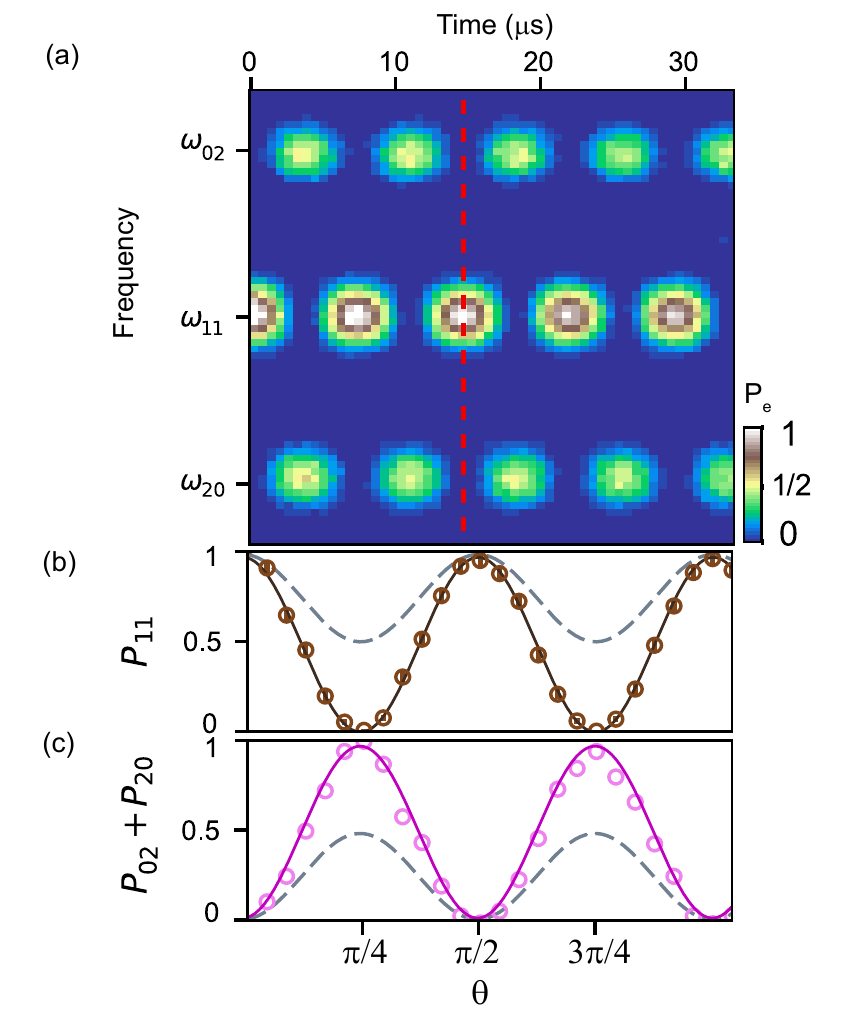}
\caption{\textbf{HOM interference between Alice and Bob.} (a) Data show the joint population as the initial state $|1,1\rangle_{\mathrm{AB}}$ evolves under $\hat{U}_{\mathrm{BS}}$. We observe out-of-phase oscillations between $P_{11}$ and an equal superposition of $|0,2\rangle_\mathrm{AB}$ and $|2,0\rangle_\mathrm{AB}$. (b), (c) 1D cuts to show the behavior of $P_{11}$ (brown), and $P_{02}+P_{20}$ (magenta) up to $\sim$ 15\,$\mu$s (red dashed line). Simulation of two indistinguishable photons is shown in solid lines and that of two distinguishable particles are shown in dashed grey lines.}
\label{fig:hom}
\end{figure}

\begin{figure*}[!t]
\centering
\includegraphics[scale=1]{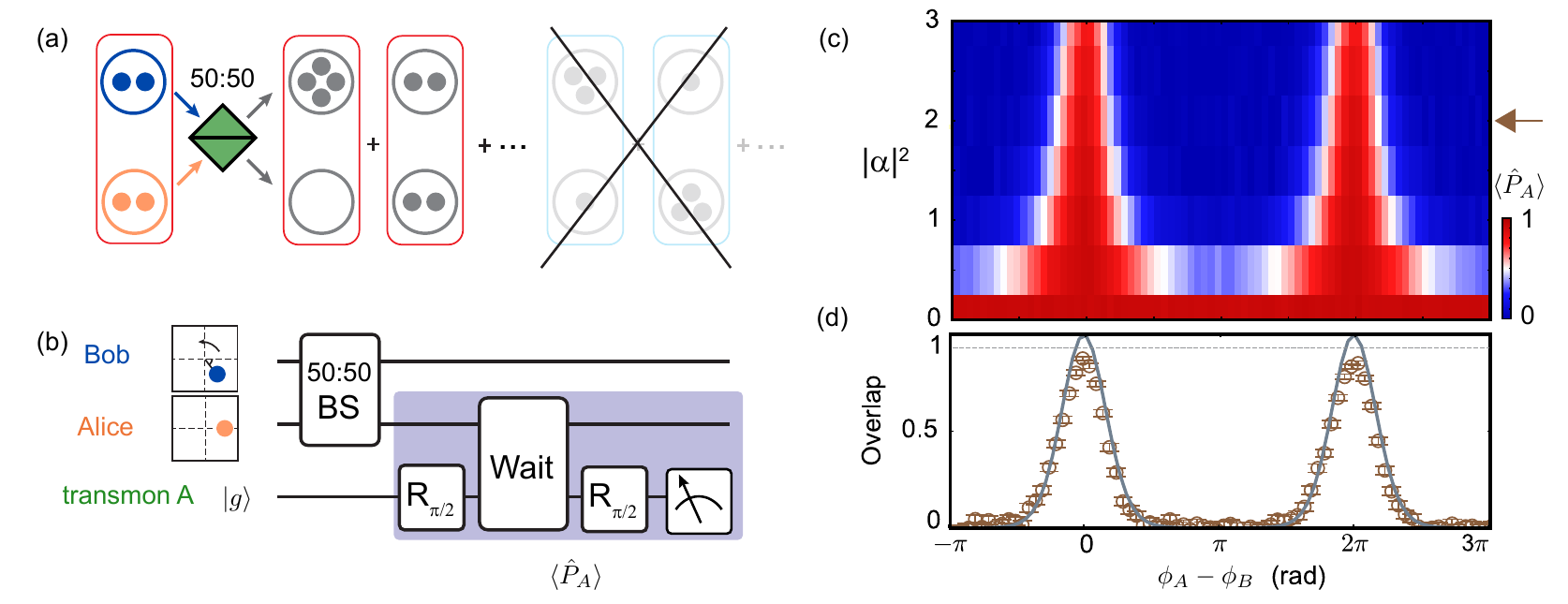}
\caption{\textbf{Measurement of quantum state overlap.} (a) When two identical bosonic systems interfere at a 50:50 BS, only even photon number outcomes are allowed due to the full destructive interference of the odd distributions. This allows us to infer the state overlap, $\mathrm{Tr}(\rho_{A}\rho_{B})$, from photon number parity measurement of one of the output ports. (b) The experimental protocol for implementing the overlap measurement between Alice and Bob. (c) Measurement of overlap between $|\psi\rangle_{A} = e^{i\phi_{A}}|\alpha\rangle$ and $|\psi\rangle_{B} = e^{i\phi_{B}}|\alpha\rangle$ as a function of the displacement angle $\phi$ and amplitude $\alpha$. Maximum overlap for each displacement is measured when $\phi_{B} = \phi_{A}$ (mod 2$\pi$). (d) 1D cut at displacement of $|\alpha|^2 \approx 2$. Data are shown in brown circles, which shows good agreement with the overlap determined by full Wigner tomography of each mode, shown in solid grey line. The dashed grey line indicates the predicated peak contrast after taking into account self-Kerr and the decoherence of the cavities.}
\label{fig:overlap}
\end{figure*}

The quality of the BS operation can also be characterized by the contrast of HOM interference, which is the hallmark of interference between two indistinguishable bosonic modes. We demonstrate that this behavior can be observed between two photons at different frequencies via the engineered frequency-converting coupling, similar to the the results described in Ref.~\cite{kobayashi2016}. We start by preparing $|1, 1\rangle_{\mathrm{AB}}$ and monitor the joint population of Alice and Bob after applying the drive tones. The photon number distribution of Alice and Bob is probed by performing a photon-number selective $\pi$ pulse on qC at different frequencies. The population of particular joint photon number state $|n, m\rangle_{\mathrm{AB}}$ is given by the probability of exciting qC at the $\omega_{nm} = \omega_{c} - n\chi_{ac} - m\chi_{bc}$ as shown in Fig.~\ref{fig:hom}(a). Two important observations can be made from this measurement. 

First, the data indicate that the engineered BS operation indeed allows two detuned photons to behave as indistinguishable particles. This is shown by near complete destruction of the $|1,1\rangle_{\mathrm{AB}}$ signal after the BS.  At this point, we also measure an equal probability of finding the system in states $|2, 0\rangle_\mathrm{AB}$ and $|0, 2\rangle_\mathrm{AB}$. This interference is a intrinsically quantum mechanical phenomena. It typically relies on the two initial photons being fully indistinguishable such that the probability amplitudes of $|1,1\rangle_{\mathrm{AB}}$ after the BS destructively interfere. When two photons are distinguishable, the classical probability distribution is observed and the measured $P_{11}$ always remain above 0.5. In this case, since the two cavity modes are far-detuned from each other, the initial excitations are completely distinct. The observation of quantum interference depends crucially on the frequency-converting coupling that can fully compensate for the energy difference between the two initial photons. Therefore, the measured HOM contrast of $98\pm 1\%$ is a direct indication of the quality of the engineered BS operation. 

Second, we know that the BS operation preserves the phase coherence of the superposition states. This is demonstrated by the near unit probability of $|1,1\rangle_{\mathrm{AB}}$ after the second BS with full extinction of $|2,0\rangle_\mathrm{AB}$ and $|0, 2\rangle_\mathrm{AB}$. Thus, we infer that the system is indeed in a coherent superposition of $|2, 0\rangle_\mathrm{AB}$ and $|0,2\rangle_\mathrm{AB}$ after the first BS because a statistically mixed state would not allow the full constructive interference of $|1, 1\rangle_{\mathrm{AB}}$. Consequently, it would lead to a reduction the $P_{11}$ contrast.  

The HOM experiment reveals an intrinsic property of bosonic systems: when two identical quantum states interfere through a 50:50 BS, the photon number parity of the output ports is always even because the odd outcomes interfere destructively, as illustrated in Fig.~\ref{fig:overlap}(a). In fact, it has been proven that the average parity measured on one of the output ports after a BS is a direct probe of the overlap between the two initial states, i.e $\langle \hat{P}_{A} \rangle = \langle \hat{P}_{B} \rangle = \mathrm{Tr}(\rho_{A}\rho_{B})$~\cite{islam2015, daley2012, ekert2002, filip2002}. This establishes a mapping between the state overlap and a single observable that is the photon number parity of one of the output modes regardless of the input states.

As a demonstration, we perform direct overlap measurement between two coherent states $|\psi\rangle_A = e^{i\phi_{A}}|\alpha\rangle$ and $|\psi\rangle_B = e^{i\phi_{B}}|\alpha\rangle$ using a single BS and robust parity measurement enabled by the natural dispersive coupling between qA and Alice in our cQED system~\cite{bertet2002, sun2014}. The experimental sequence is outlined in Fig.~\ref{fig:overlap}(b), where $\phi_{A}$ is fixed and the parity of Alice is measured as a function of $\phi_{B}$ for variable displacement amplitude $\alpha$ in both cavities. As expected, the maximum overlap is measured when $\phi_{A} = \phi_{B} (\mathrm{mod}\, 2\pi),$ and as the displacement increases, the measured $\langle \hat{P}_{A}\rangle$ becomes more sharply peaked. This is in good agreement with the ideal $\mathrm{Tr}(\rho_{A}\rho_{B})$ calculated using simulated full Wigner functions of each mode. We do observe a reduction in the overall contrast at higher photon numbers which can be accounted for by known imperfections of the parity measurement and the BS operation~\cite{supplement}. 

\begin{figure}[!t]
\centering
\includegraphics[scale=1]{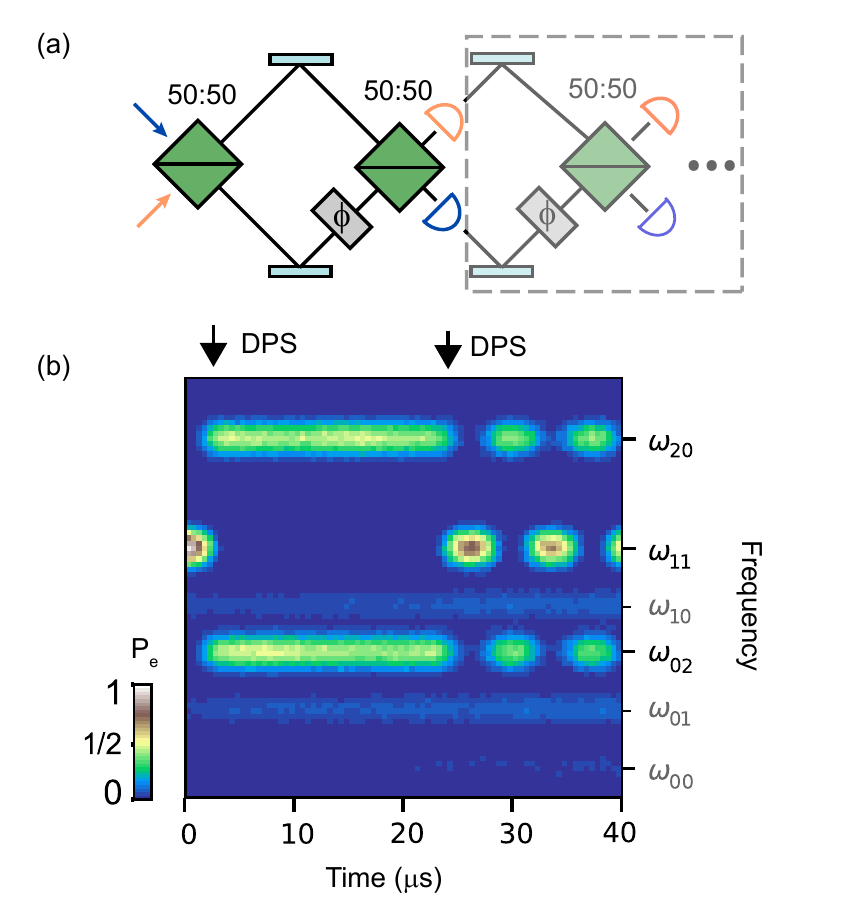}
\caption{ \textbf{Cascaded Mach-Zehnder interferometers composed of BS, DPS and photon counters.} (a) Conceptual implementation of programmable cascaded Mach-Zehnder (MZ) interferometers between two detuned modes. We can measure the state of the system after each operation and implement DPS on-demand. (b) Evolution of $|1, 1\rangle_\mathrm{AB}$ in cascaded microwave MZ interferometers, measured by sweeping the frequency of a photon number selective $\pi$ pulse on qC and the duration of the drive tones. After the first BS, a coherent superposition of $|0, 2\rangle_\mathrm{AB}$ and $|2, 0\rangle_\mathrm{AB}$ ($|\Psi\rangle$) is created. Subsequently, we impart a differential phase $\phi = \pi$, which leaves the system in $|\Phi\rangle$ and causes $|1, 1\rangle_\mathrm{AB}$ to destructively interfere through the subsequent BS's. We then introduce another differential $\pi$-phase such that the system is brought back the initial superposition $|\Psi\rangle$ which re-focuses to $|1, 1\rangle_\mathrm{AB}$ at the following BS as in the standard HOM interference. }
\label{fig:mz}
\end{figure}

In addition to the engineered BS operation, we can also implement on-demand DPS to Alice via its natural dispersive coupling ($\chi_1$) to qA. This is governed by the unitary $\hat{U}_{\mathrm{DPS}}(t) = |g\rangle\langle g|\otimes \hat{I} +|e\rangle\langle e|\otimes e^{i{\phi}\hat{a}^{\dagger}\hat{a}}$, where $\phi = \chi_{\mathrm{1}}t$. This implementation has two major advantages. It is fully programmable: the resulting differential phase, $\phi \in [0, 2\pi]$, is simply controlled by the evolution time $t$, which can be tuned on the fly. Furthermore, it is photon-number independent. $\hat{U}_{\mathrm{DPS}}$ allows us to impart the same phase to each individual photon in Alice. Therefore, it is naturally compatible with more complex interference experiments involving multi-photon states. 

Combining the BS and DPS capabilities, we can construct cascaded MZ interferometers and program them to perform different interference experiments on the fly (Fig.~\ref{fig:mz}(a)).  As a simple example, we initialize the system in $|1, 1\rangle_{\mathrm{AB}}$. After a single BS, the system reaches the superposition state $|\Psi\rangle = \frac{1}{\sqrt{2}}(|0, 2\rangle_\mathrm{AB} + e^{i\varphi}|2, 0\rangle_\mathrm{AB})$. Subsequently, we can impart a phase on Alice by exciting qA for a time $\pi/\chi_{1} \sim 500\,$ns. This operation changes the relative phase between Alice and Bob, leaving the system in the state $|\Phi\rangle = \frac{1}{\sqrt{2}}(|0, 2\rangle_\mathrm{AB}+ e^{i(\varphi + \pi)}|2, 0\rangle_\mathrm{AB})$. This is now a `dark state' of $\hat{U}_{\mathrm{BS}}$ because the probability amplitudes of $|1, 1\rangle_{\mathrm{AB}}$ always interfere destructively, forcing the system to remain in $|\Phi\rangle$ through all subsequent beamsplitters. We then bring the system back to $|\Psi\rangle$ and recover the HOM-type interference by implementing a second DPS of $\pi$ on Alice. The reduced contrast in the revival of $|1, 1\rangle_{\mathrm{AB}}$ can be attributed to the non-idealities due to both the self-Kerr and the dephasing of the memory modes during the evolution in the dark state.

Such cascaded interferometers, similar to optical implementations, such as in Ref.~\cite{crespi2013}, offers a versatile and scalable scheme to study complex interference phenomena and boson statistics. In particular, all components in our implementation are by design compatible with multi-photon states. Combining this with our ability to prepare complex bosonic states and efficiently probe their statistics using the transmon ancilla, we can easily extend such cascaded interferometers to investigate the interference between a larger number of excitations. A simple demonstration is shown in Fig.~S5, where  memories initialized in a state containing three excitations interference through a series of engineered BS~\cite{supplement}.  

\section{IV. Conclusion}
In conclusion, we have engineered  a robust BS operation between bosonic quantum memories in a fixed-frequency cQED architecture. With this, we demonstrated high-contrast HOM interference between two memories at different frequencies. Furthermore, we combined this coupling with single-cavity phase control and photon number parity measurement to implement efficient overlap measurements and on-demand manipulation of the signature of two-photon interference. Taking advantage of the tunability of the BS and our ability to implement DPS on-the-fly, we constructed highly programmable cascaded interferometers capable of manipulating interference statistics on demand. Our implementation can be directly extended to higher photon numbers to study more complex interference between multi-photon states~\cite{supplement}. The robust BS and DPS operations demonstrated in this work form an essential toolset for implementing gates between logical qubits encoded in superconducting cavities~\cite{mirrahimi2014, lau2016, Lloyd2014}. Additionally, our results also provide the essential components required for implementing LOQC protocols~\cite{knill2001, chirolli2010} in the cQED framework. 

\section*{Acknowledgements}
We thank M. H. Devoret for the helpful discussions; N. Frattini, K. Sliwa, M.J. Hatridge and A. Narla for providing the Josephson Parametric Converter (JPC); N. Ofek and P. Reinhold for providing the logic and control interface for the field programmable gate array (FPGA) used in of this experiment. This research was supported by the U.S. Army Research Office (W911NF-14-1-0011). Y.Y.G. was supported by an A*STAR NSS Fellowship; B.J.L. is supported by Yale QIMP Fellowship; S.M.G. by the National Science Foundation (DMR-1609326); L.J. by the Alfred P. Sloan Foundation (BR 2013-049) and the Packard Foundation (2013-39273). Facilities use was supported by the Yale Institute for Nanoscience and Quantum Engineering (YINQE), the Yale SEAS cleanroom, and the National Science Foundation (MRSECDMR-1119826).

\bibliography{../ref_lib.bib}        

\end{document}


\title{Supplementary materials:\\
Programmable interference between two microwave quantum memories }
\author{Yvonne Y. Gao}
\email[corresponding author:]{yvonne.gao@yale.edu}
\affiliation{Departments of Physics and Applied Physics, Yale University, New Haven, Connecticut 06511, USA}
\affiliation{Yale Quantum Institute, Yale University, New Haven, Connecticut 06511, USA}
\author{B.J. Lester}
\affiliation{Departments of Physics and Applied Physics, Yale University, New Haven, Connecticut 06511, USA}
\affiliation{Yale Quantum Institute, Yale University, New Haven, Connecticut 06511, USA}
\author{Yaxing Zhang}
\affiliation{Departments of Physics and Applied Physics, Yale University, New Haven, Connecticut 06511, USA}
\affiliation{Yale Quantum Institute, Yale University, New Haven, Connecticut 06511, USA}
\author{C. Wang}
\affiliation{University of Massachusetts, Amherst, MA 01003-9337 USA}
\author{S. Rosenblum}
\affiliation{Departments of Physics and Applied Physics, Yale University, New Haven, Connecticut 06511, USA}
\affiliation{Yale Quantum Institute, Yale University, New Haven, Connecticut 06511, USA}
\author{\\L. Frunzio}
\affiliation{Departments of Physics and Applied Physics, Yale University, New Haven, Connecticut 06511, USA}
\affiliation{Yale Quantum Institute, Yale University, New Haven, Connecticut 06511, USA}
\author{Liang Jiang}
\affiliation{Departments of Physics and Applied Physics, Yale University, New Haven, Connecticut 06511, USA}
\affiliation{Yale Quantum Institute, Yale University, New Haven, Connecticut 06511, USA}
\author{S.M. Girvin}
\affiliation{Departments of Physics and Applied Physics, Yale University, New Haven, Connecticut 06511, USA}
\affiliation{Yale Quantum Institute, Yale University, New Haven, Connecticut 06511, USA}
\author{R.J. Schoelkopf}
\affiliation{Departments of Physics and Applied Physics, Yale University, New Haven, Connecticut 06511, USA}
\affiliation{Yale Quantum Institute, Yale University, New Haven, Connecticut 06511, USA}
\date{\today}

\pacs{}
\maketitle

\section{Device architecture and system parameters}
\begin{figure}[!b]
\centering
\includegraphics[scale=0.75]{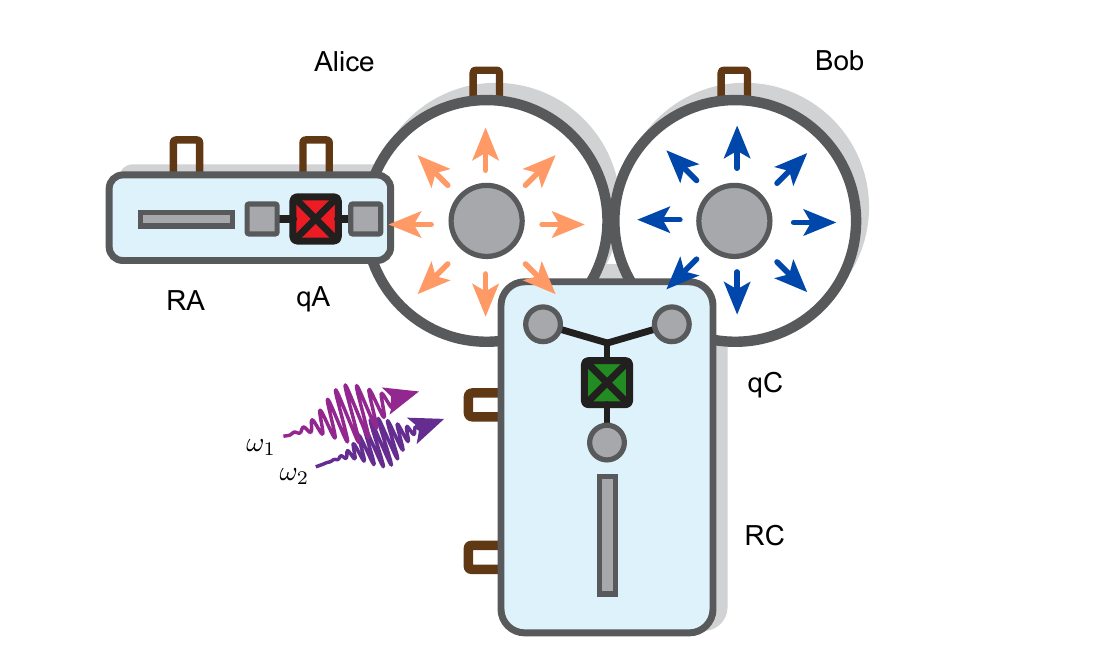}
\caption{\textbf{A cartoon showing the top view of the 3D double-cavity cQED system}. The center transmon ancilla provides nonlinear coupling between Alice and Bob. The package can accommodate two additional transmon ancillae. In this experiment, only one (qA) is included in the device. The RF drives are coupled to the system through the drive port of qC.}
\label{sfig:cartoon}
\end{figure}

Our cQED system includes two three dimensional (3D) superconducting microwave cavities, Alice and Bob, two quasi-planar readout resonators, RA and RC, and two transmon-type superconducting qubits, qA and qC. All components are housed in a  single block of high-purity (4N) aluminum in the structure shown in Fig.~\ref{sfig:cartoon}. Alice and Bob act as quantum memories that are capable of coherently storing quantum information in bosonic states. They are formed by 3D coaxial transmission lines that are short-circuited on one end and open-circuited on the other by virtue of a narrow circular waveguide \cite{reagor2016}. The resonance frequency of the cavities' fundamental modes are determined by the lengths of the transmission lines (4.8 and 5.6 mm respectively for Alice and Bob).  An elliptical tunnel is machined between Alice and Bob, allowing the insertion of a chip containing the transmon ancilla and readout resonator into the cavities.  Two additional tunnels are machined on either side of Alice and Bob to allow the incorporation of additional transmon and readout channels. Each mode is coupled to the fridge input/output lines via standard SMA couplers. 

The whole package is chemically etched after machining to improve the surface quality of the cavities. The superconducting transmons are fabricated on sapphire substrates using electron-beam lithography and the standard shadow-mask evaporation of Al/AlOx/Al Josephson junction. During the same fabrication process, a separate strip of the tri-layer film is also deposited. Together with the wall of the tunnel, it forms a hybrid planar-3D $\lambda$/2 stripline resonator that is capacitively coupled to the transmon.  This design combines the advantages of both precise, lithographic control of critical dimensions and the low surface/radiation loss of 3D structures \cite{axline2016}. The chip containing these structures is inserted into the tunnel such that the transmon antenna(a) protrudes into the cavities to the desired capacitive coupling to Alice and Bob.

This system is an extension of the device used in Ref.~\cite{wang2016}. The additional tunnels allow individual ancillae, accompanied by their respective readout resonator, to couple to each cavity in order to provide fast, independent cavity manipulations and tomography. Only one additional ancilla (qA) is required for this particular experiment. It couples to only Alice and is used perform differential phase shifts (DPS) in the Mach-Zehnder interometer described in the main text. The parameters of all relevant components are summarized in Table~\ref{table:params_multi_ancillae}. 

\begin{table*}[!htb]
\centering  
\vspace{2ex}
\begin{tabular}{c c c c c c } 
\hline\hline\\[-2ex]
		&\,\,\, Frequency\,\,\,\, 	& 	Nonlinear &interactions:	 \\
		& \,\,\, $\omega/2\pi$\,\,\,\,	& Alice			& Bob			& qA 		 & qC 		 \\
\hline\\[-2ex]
Alice\,	  &5.554\,GHz\,	& 4\,kHz\,			&  $\lesssim$1\,kHz\,	 & 1.01\,MHz\,\,\,	&0.62\,MHz\, 	\\
Bob\,		 & 6.543\,GHz\,	& $\lesssim$1\,kHz\,	& 2\,kHz\,			& ($\sim$ 0)\,\,\, 	&0.26\,MHz\,	\\
qA\, 		& 4.989\,GHz\,	& 1.01\,MHz\,		&  ($\sim$ 0)\,		& 185\,MHz\,\,\,		& N.A\,		\\
qC\,           & 5.901\,GHz\,	& 0.62\,MHz\,		&0.26\,MHz\, 		& N.A\,\,			& 74\,MHz\,	\\
RA\,		& 7.724\,GHz\,	& (4\,kHz)\,		& ($\sim$ 0)\,		& 1.2\,MHz\,\, \,		&($\sim$ 0)\, 	\\
RC\,		& 8.062\,GHz\,	& ($\sim$ 2\,kHz)\,	& ($\sim$ 1\,kHz)\,		& ($\sim$ 0)\,\,\,	&1.1\,MHz\, 	\\[0.5ex]
\hline
\end{tabular}
\caption{\textbf{Hamiltonian parameters of all cQED components.} Values that are within a parenthesis are estimated/simulated parameters. Some non-linear couplings, such as $\chi$ between qA and Bob, are omitted because they are too small to be simulated and measured.}\label{table:params_multi_ancillae}
\end{table*}

We highlight the large detuning between Alice and Bob, which results in a small cross-Kerr ($\chi_{ab} \lesssim$1\,kHz) between them. This ensures that there is minimal unwanted interactions between the two modes in the absence of the RF drives, therefore allowing the operation to have a large on-off ratio. Further, it is also important that both Alice and Bob have relatively weak self-Kerr, which is the non-linearity inherited from their couplings to the ancillae. Large self-Kerr would cause imperfections in the interference between Alice and Bob, especially at larger photon numbers. Finally, we comment on the choice of the resonance frequency of qC. It is placed in between the frequencies of Alice and Bob so that it has comparable, but different, dispersive coupling strength to each cavity. This allows us to use qC as a meter to efficiently probe the joint photon number state, $|n, m\rangle_{AB}$, in Alice and Bob by measuring its transition frequency $\omega_{ge}$, which is given by $\omega_{\mathrm{ge}} = \omega^{00}_{\mathrm{ge}} - n_{A}\chi_{\mathrm{ac}} - m_{B}\chi_{\mathrm{bc}}$. 

We characterize the coherence of each component in the system using standard cQED measurements. The results are summarized in Table~\ref{table:t1t2s}. 
\begin{table}[!htb]
\centering  
\vspace{2ex}
\begin{tabular}{c c c c c} 
\hline\hline\\[-2ex]
		& T$_{1}$ ($\mu$s)\,\,	&  T$_{2}$ ($\mu$s)\,\, &  T$_{2E}$ ($\mu$s)\,\,  & Population\\
\hline\\[-2ex]
Alice\,	         & 400-500\,\,\, 		& 400-600\,\,\,		& N.A\,\,	& $<$1\% \\
Bob\,		& 400-500\,\,\,		&400-600\,\,\,		& N.A\,\,	& $<$1\%\\
qA\, 		& 70\,\,\,		& 15-20\,\,	\,	& 30-40\,\,	 & 2-3\%\\
qC\,          & 50\,\,\,		& 10-15\,\,	\,	& 25-40\,\,    & 2-3\% \\
\hline
\end{tabular}
\caption{\textbf{Coherence properties of the the system.} The device exhibits some fluctuations in its coherence times. In particular, the relatively low $T_2$ of qA and qC are likely a result of low-frequency mechanical vibrations.}\label{table:t1t2s}
\end{table}

\section{The driven Josephson circuit Hamiltonian}
Similar to the treatment in~\cite{leghtas2015}, the full Hamiltonian describing the system consisting of Alice, Bob, qC, and two RF drive tones applied to qC can be written as
\begin{align}\label{eq:hamiltonian}
\hat{H}/\hbar = &\omega_{a}\hat{a}^{\dagger}\hat{a} +  \omega_{b}\hat{b}^{\dagger}\hat{b} +  \omega_{c}\hat{c}^{\dagger}\hat{c} - \frac{E_{\mathrm{J}}}{\hbar}(\cos{\hat{\varphi}} + \frac{\hat{\varphi}^2}{2}) \nonumber\\
&+ 2\mathrm{Re}[\epsilon_{1}e^{-i\omega_{\mathrm{1}}t} + \epsilon_{2}e^{-i\omega_{\mathrm{2}}t}](\hat{c}^{\dagger} + \hat{c})
\end{align}
where $\omega_{k}$ is the frequency of each mode, $k$, and $\hat{\varphi}$ is the flux across the junction, which can be decomposed into a linear combination of the phase across each mode:
\begin{equation}
\hat{\varphi} = \sum_{k=a,b,c} {\phi}_{k}(\hat{k}^{\dagger}+\hat{k}) 
\end{equation}
This Hamiltonian captures the behavior of the system when irradiated by two drives with complex amplitudes, $\epsilon_{1}$ and $\epsilon_{2}$, and frequencies  $\omega_{1}$ and $\omega_{2}$, respectively. In our particular configuration, the drive tones are predominantly coupled to qC, as assumed in Eq.~\ref{eq:hamiltonian}.  Further, we assume that the drive tones are stiff, i.e the annihilation and creation of a photon from the drive does not cause any change to the mode. Therefore, they can simply be treated as classical drives. 


We eliminate the fastest time scales corresponding to the resonance frequencies of each mode using the following unitary transformation
\begin{equation}
\hat{U} = e^{-i\omega_{\mathrm{ge}}t\hat{c}^{\dagger}\hat{c}}e^{-i\omega_{a}t\hat{a}^{\dagger}\hat{a}}e^{-i\omega_{b}t\hat{b}^{\dagger}\hat{b}}
\end{equation}
Then we make a displacement transformation for qC such that $\hat{c}\rightarrow \hat{c} + \xi_1 e^{-i\omega_1 t}+\xi_2 e^{-i\omega_2 t}$. 
In this new frame, we express the amplitudes of the drives, $\xi_{1(2)}$, as a function of the amplitudes of the drive tones and their respective detunings from the $|g\rangle-|e\rangle$ transition frequency of qC, $\omega_{\mathrm{ge}}$: 
\begin{equation}
\xi_{1(2)} = -\frac{i \epsilon_{1(2)}}{(\tilde{\kappa}/2 + i (\omega_{\mathrm{ge}} - \omega_{1(2)})}
\end{equation}
where $\tilde{\kappa}$ is the effective decay rate of the mode to which the drives couple to primarily. In this case, it is the decay associated with qC, which is at least an order of magnitude smaller than the coupling rates associated with the driven interaction.

Now, we derive the effective Hamiltonian by expanding the cosine potential in Eq. (1) to the 4th order and perform the standard rotating wave approximations (RWA). As the frequency matching condition is satisfied when $\omega_2 - \omega_1 = \omega_b - \omega_a$, the only 4th order, non-rotating terms are: 
\begin{align}
\hat{H} &= \hat{H}^{1(2)}_{ss} + \hat{H}_{\mathrm{Kerr}} + \hat{H}_{\mathrm{int}}\\
\hat{H}_{\mathrm{int}} &=  -E_{\mathrm{J}}\phi^2_{c}\phi_{a} \phi_{b}(\xi_{1}\xi_2^{*}\hat{a}^{\dagger}\hat{b} + \xi_{1}^{*}\xi_2\hat{a}\hat{b}^{\dagger}) \\
\hat{H}^{1(2)}_{\mathrm{ss}}&\approx -E_{\mathrm{J}}\phi^4_{c} |\xi_{1(2)}|^{2}\hat{c}^{\dagger}\hat{c} = -2\alpha|\xi_{1(2)}|^{2}\hat{c}^{\dagger}\hat{c}\\
\hat{H}_{\mathrm{Kerr}} & = -\sum_{k=a,b,c} \frac{E_{\mathrm{J}}{\phi}_{k}^4}{4}\hat{k}^{\dagger}\hat{k}^{\dagger}\hat{k}\hat{k} -  E_{\mathrm{J}}{\phi}_{a}^2\phi_{b}^2\hat{a}^{\dagger}\hat{a}\hat{b}^{\dagger}\hat{b} \nonumber\\
&\quad - E_{\mathrm{J}}{\phi}_{a}^2\phi_{c}^2\hat{a}^{\dagger}\hat{a}\hat{c}^{\dagger}\hat{c}- E_{\mathrm{J}}{\phi}_{b}^2\phi_{c}^2\hat{b}^{\dagger}\hat{b}\hat{c}^{\dagger}\hat{c}
\end{align}
where $H_{\mathrm{int}}$ is the desired interaction term. $\hat{H}_{\mathrm{Kerr}}$ describes the self-Kerr and cross-Kerr coupling terms~\cite{nigg2012}, which do not depend on the RF drives. They are calibrated independently using methods developed in Ref.~\cite{kirchmair2013}. Finally, $H^{1(2)}_{\mathrm{ss}}$ captures the Stark shift of the resonance frequency of qC in the presence of each RF drive. In the limit of $|\omega_{1,2} - \omega_{\mathrm{ge}}|\gg\alpha$, the Stark shift only depends on the normalized drive strengths, $\xi_1$ and $\xi_2$, as well as $\alpha = \frac{1}{2}E_{\mathrm{J}}\phi^4_{c}$ (Eq.~S6). For the drive configuration we have chosen in this experiment, the detuning between $\omega_1$ and $\omega_{\mathrm{ge}}$ is only a factor of $\sim$2 larger than $\alpha$. Thus, we must take into account the other slowly rotating terms in the cosine expansion which have the form $\alpha\xi_{1}e^{-i\delta t}(\hat{c}^{\dagger})^2\hat{c}$ + h.c, where $\delta = \omega_1 - \omega_{\mathrm{ge}}\approx157\,$MHz. We treat this perturbatively and to the leading order in the drive power, this term results in a modification factor $\delta/(\delta + \alpha)$ to the Stark shift which must be take into account in our calibration process. Further, such non-resonant terms also introduce a correction factor to the coupling strength $\tilde{g} =g(1+ 2\alpha/(\delta + (\omega_2 - \omega_\mathrm{ge}) + (\omega_a-\omega_{\mathrm{ge}}) + (\omega_b - \omega_{\mathrm{ge}})))^{-1}$~\cite{zhang2018}. We measure $\alpha$ independently using simple spectroscopic methods. It is also worth noting that the drives are applied adiabatically with a smooth ring-up and ring down time of $\sim 100$\,ns to ensure that no additional spectral components are present in the drive tones. 

\begin{figure}[!h]
\centering
\includegraphics[scale=1]{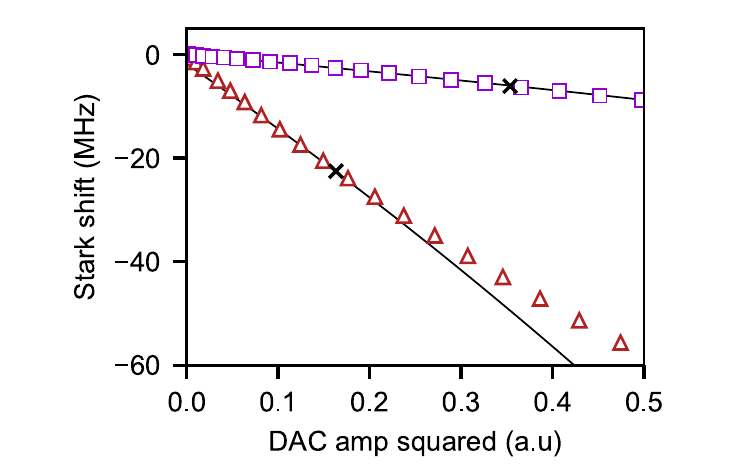}
\caption{\textbf{Stark shift due to each drive tone.} Data show the shift of qC resonance frequency as a function of the input power of two drives at $\omega_1$ (red) and $\omega_2$ (magenta) respectively. The drive tone at $\omega_1= \omega_{\mathrm{ge}} + 157\,$MHz results in more significant shifts of $\omega_{\mathrm{ge}}$ due to the small detuning. In contrast the much further detuned with $\omega_2= \omega_{\mathrm{ge}} + 1.148\,$GHz leads to a much smaller shift. Solid lines are linear fits to the data based on the derivations in Eq. S6. Black crosses indicate the DAC amp squared used in the interference experiments presented the main text.}
\label{sfig:pump_cal}
\end{figure}

Due to the difference in the detuning to qC from each RF drive, the resulting Stark shifts differ quite significantly in this configuration (Fig.~\ref{sfig:pump_cal}). For the further detuned drive (magenta), the Stark shift exhibits linear dependence on the drive power, in good agreement with the model using 4th order expansion. However, the drive tone that is spectrally much closer to qC, the resulting Stark shift deviates from the predicted linear trend at higher powers. Such deviation becomes strong when the Stark shift becomes comparable to the detuning of the drive from $\omega_{\mathrm{ge}}$. We verified that the observed AC Stark shift is well captured by our model that keeps up to quartic terms in the expansion of cosine potential in Eq. (1) and treats the drives non-perturbatively; see Ref.[7]. Here, by using the full DAC range of our RF input signal, we can tune the effective drive ampliudes up to $\xi_1 \approx 0.75$ and $\xi_2 \approx 0.25$. In practice, we operate at where the dependence on the drive power is roughly linear (black crosses) and the effects of six order terms are negligible. From this calibration, the effective coupling strengths, $g$, are computed for a chosen set of $\xi_{1(2)}$ from the single photon dynamics under $U_{\mathrm{BS}}$ and compared against the theory prediction as shown in Fig.\,2(c) of the main text.  

The particular choice of drive frequencies is used in order to avoid exciting any higher order transitions of qC. In principle, we could move the frequencies of both drives together such that the two drives are roughly equally detuned from qC. However, this requires one tone to be placed below the $\omega_{\mathrm{ge}}$, which makes the spectral overlap with the $|e\rangle-|f\rangle$ or $|f\rangle-|h\rangle$ transitions more likely. 

\section{drive-assisted absorption of cavity photons}\label{loss}
The transmon, qC, that supplies the non-linearity needed for the controlled beamsplitter operation between cavities typically has much shorter lifetime than the cavities. As a result, hybridization of the cavities with the qC leads to a shorter cavity lifetime. To leading order in the transmon-cavity coupling, the rate of cavity  decay inherited from the qC via the ``inverse Purcell effect" \cite{reagor2016} is given by Fermi's golden rule:
\begin{equation}
\kappa_\gamma = -2  |\lambda|^2 \Im  \frac{1}{(\omega_\mathrm{cav}-\omega_\mathrm{ge}) + i\gamma/2},
\end{equation}
where $\lambda$ is the coupling strength between the cavity with frequency $\omega_\mathrm{cav}$ and the transmon qubit. For large detuning between transmon and the cavity $|\omega_\mathrm{cav}-\omega_\mathrm{ge}|\gg \gamma$, we have $\kappa_\gamma \approx |\lambda/(\omega_\mathrm{cav}-\omega_\mathrm{ge})|^2 \gamma$. This decay rate can be thought of as the result of hybridization between a cavity excitation (photon) and an excitation of the transmon. Such inherited decay can be largely suppressed by using a large $|\omega_\mathrm{cav}-\omega_\mathrm{ge}|$, as is the case in the present experiment.

\begin{figure}[!h]
\centering
\includegraphics[scale=1]{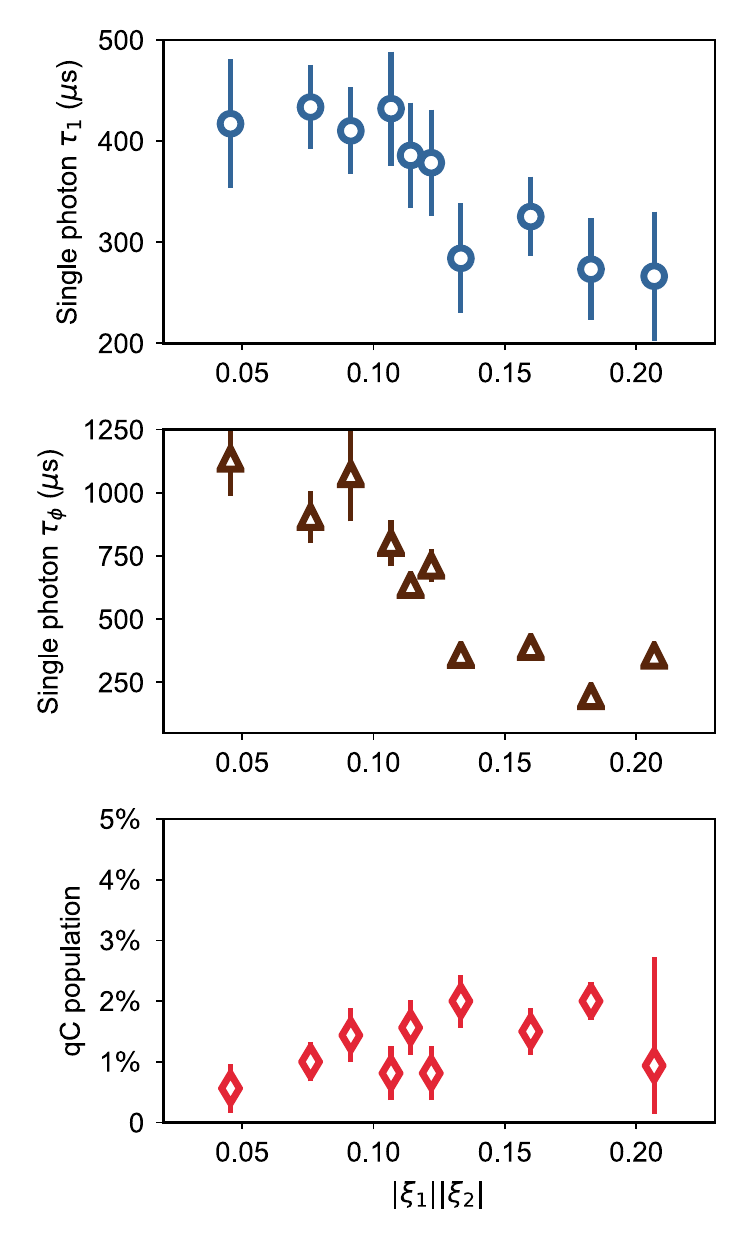}
\caption{\textbf{The extracted coherence properties of a single excitation under $\hat{U}_{\mathrm{BS}}$.} (a) The decay constant of the $P_{10} + P_{01}$ extracted from a single exponential fit. (b) The dephasing times extracted from $P_{10}/(P_{10} + P_{01})$. (c) The independently measured population of qC after a single BS operation}
\label{sfig:bs_decays}
\end{figure}

In the presence of the RF drives, a cavity photon combined with drive photons can also hybridize the excitations of qC from the ground state to higher excited states due to multi-photon resonances. Taking into account these processes yields a generalized formula for $\kappa_\gamma$
\begin{align}\label{eq:kappa_gamma}
\kappa_\gamma = -2 \sum_{K K'}|M_{K K'}|^2 \Im \frac{1}{(\omega_\mathrm{cav}-\nu_{KK'})+i\gamma_{KK'}/2}.
\end{align}
Each term in the summation refers to a process in which one cavity photon, $K$ photons from the low frequency drive and $K'$ photons from the high frequency one, excite qC from the ground state to the $(K+K'+1)$-th excited state. Hence, the resonance frequency $\nu_{KK'}$ is given by the relation \[ \nu_{K K'} + K\omega_\mathrm1 + K'\omega_\mathrm2 = \mathcal E_{K+K'+1}-\mathcal E_0, \] where $\mathcal E_K$ is the AC-Stark-shifted energy of the $K$-th excited state of the transmon. $M_{KK'}$ and $\gamma_{KK'}$ are the matrix element and width of the same resonance process, respectively. In the weak drive regime, we have from perturbation theory that \[ |M_{KK'}|\propto|\xi_1|^{|K|}|\xi_2|^{|K'|},\]\[ \gamma_{KK'}\approx (K+K'+1)\gamma.\] To go beyond the weak drives,  we have developed a Floquet theory to study the dynamics of the driven non-linear transmon which will be described in depth in a separate theory analysis~\cite{zhang2018}.


Experimentally, we observe a consistent degradation of the BS quality as the drive powers become higher. To investigate qualitatively, we extract the coherence time scales of a single excitation evolving under the engineered $\hat{U}_{\mathrm{BS}}$ at each $|\xi_1||\xi_2|$ as described in the main text. As we sweep the drive powers in each of these experiments, we adjust the drive frequencies accordingly to ensure that the resonance condition is still satisfied. As show in Fig.~\ref{sfig:bs_decays}, both $\tau_{1}$ and $\tau_{\phi}$ remains basically independent of $|\xi_1||\xi_2|$ at low drive powers but a significant reduction is observed when $|\xi_1||\xi_2|\gtrsim 0.15$. This is not surprising since when the coupling is weak, the systems is essentially limited by its intrinsic decoherence time scales, i.e. cavity photon loss and dephasing. In this regime, the transmon excited levels are only virtually involved in the process and hence, its decoherence properties do not influence the quality of the operation. However, as we increase the drive powers the transmon's participation increases. More specifically, for the frequency configuration used in the experiment,participation of transmon's $|f\rangle$ state increases most strongly as drive strengths increase. This is because $\nu_{10}$ becomes increasingly close to $\omega_a$ due to AC Stark shift. In other words, the process in which one cavity photon at $\omega_a$ together with one drive photon at $\omega_1$ excites the transmon from $|g\rangle$ to $|f\rangle$ becomes less virtual. This subjects the system to the much less favorable decoherence time scales of the ancilla, which is more than an order of magnitude faster than that of cavities. It also results in a increase in the probability of finding the transmon in the excited states as shown in Fig.~\ref{sfig:bs_decays}(c).

\begin{figure}[!b]
\centering
\includegraphics[scale=1]{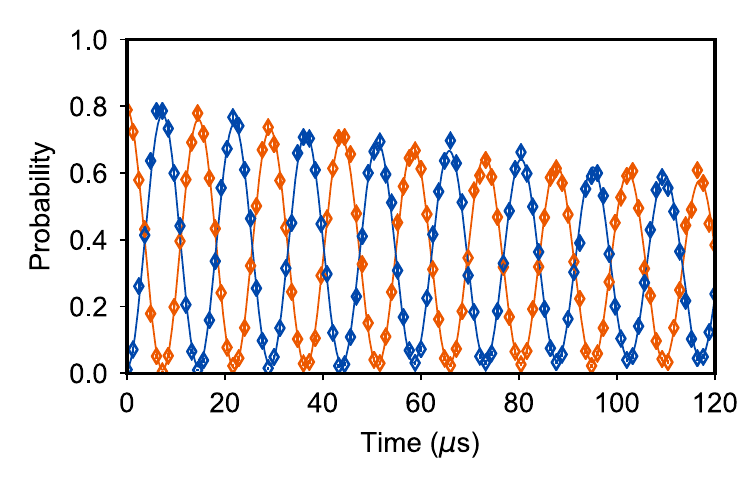}
\caption{\textbf{Oscillations of $P_{10}$ and $P_{01}$ under $\hat{U}_{\mathrm{BS}}.$} A single excitation is prepared in Alice and we monitor its evolution under the engineered unitary operation. We shown that $P_{10}$ (orange) and $P_{01}$ (blue) coherently oscillates with equal contrast and opposite phase. }
\label{sfig:p1001}
\end{figure}

\begin{figure*}
\centering
\includegraphics[scale=1]{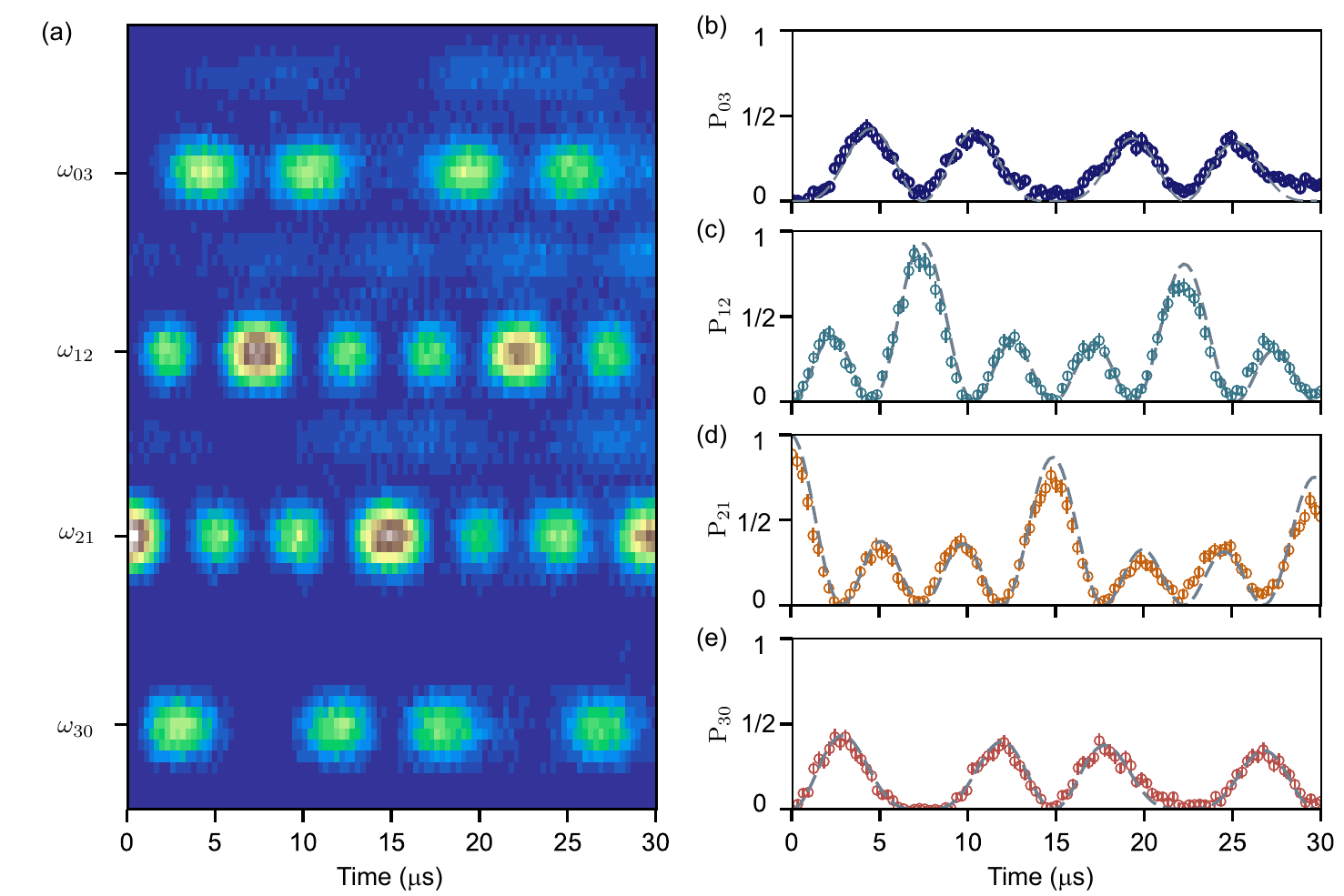}
\caption{\textbf{Interference of a multiphoton state} (a) Measurement of joint photon population distribution of the initial state  when the system is initialized in $|2, 1\rangle_{AB}$ evolving under $\hat{U}_{\mathrm{BS}}$ using qC as the meter. With a spectrally-selective pulse on qC, the probability of detecting each outcome is indicated by the probability of successively flipping qC at a particular frequency. (b)-(e) 1D cuts along the time axis for each of the possible output states. Data is shown in open circles and simulation with decoherence of the cavities taken into account is shown in grey dashed line}
\label{sfig:a2b1}
\end{figure*}

The interference experiments described in the main text are performed at $|\xi_1||\xi_2|\approx0.1$. This particular power is chosen as a compromise between minimizing qC participation and maximizing the effective coupling strength. At this drive power, we are able to perform a relatively fast BS operation ($T_{\mathrm{BS}}\lesssim$ 5\,$\mu$s $\ll \kappa_{a (b)}$) while still ensuring that qC remains largely in its ground state during the process. We show in Fig.~\ref{sfig:p1001} the oscillation of both $P_{10}$ and $P_{01}$, which have equal contrast and are exactly out of phase with each other. This implies that the excitation is strictly confined to the Hilbert space of Alice and Bob, with minimal participation of the excited levels of qC. The maximum contrast of the $P_{10}$ oscillation is $\approx$ 0.81. This is a result of the imperfections in both the state preparation and the joint photon number measurement which requires a long (4.8\,$\mu$s), spectrally-selective $\pi$ pulse on qC. These two effects can be calibrated using a Rabi experiment after preparing the initial state $|1, 0\rangle_{\mathrm{AB}}$. The measured contrast is $82\pm2\%$, consistent with that of the $P_{10}$ oscillation. In the main text, we have normalized the results of joint photon number measurements by this independently extracted scaling factor. The sum of $P_{10}$ and $P_{01}$ gives the overall decay envelope, which is consistent with the average intrinsic photon loss rate of Alice and Bob, as shown in Fig.~2(b) of the main text. 

\section{Multiphoton interference}

One advantage of using cavity states as memories is the availability of large Hilbert space.  This allows the efficient encoding of quantum information using a variety of different schemes such as the Binomial code and the cat code. To demonstrate that our engineered beamsplitter operation is compatible multiphoton states, we now perform the same type of interference studies with more excitations. 

In the first example, we initialize the system in $|2, 1\rangle_\mathrm{AB}$. Due to the absence of an independent ancilla that couples to Bob, it is rather cumbersome to prepare a Fock state in Bob directly using OCT pulses. To overcome this, we instead first initialize the system in $|1, 0\rangle_{AB}$. Subsequently, we use the engineered bilinear coupling to perform a SWAP operation and transfer this excitation from Alice to Bob. Finally, we complete the preparation by putting two photons in Alice using a numerically optimized OCT pulse. The entire process takes $\sim$ 5 $\mu$s and produces the desired state with $\sim$ 85\% fidelity with some spurious populations in $|0, 0\rangle_\mathrm{AB}$, $|1, 0\rangle_\mathrm{AB}$, and $|2, 0\rangle_\mathrm{AB}$. However, since the interference conserves total photon number and joint parity, the spurious populations do not change the statistics of the outcomes of $|2, 1\rangle_\mathrm{AB}$ undergoing $\hat{U}(\theta)$. Instead, they result in a deterministic reduction of the measurement contrast. 

The probability of each possible outcome is measured with a spectrally-selective pulse on qC as shown in Fig.~\ref{sfig:a2b1}(a). Even with only three excitations, the interference dynamics is already rather complex. As we extend our study to include more excitations, it can quickly become a non-trivial problem to predict the dynamics. This type of multi-excitation interference is often referred to as the generalized HOM interference, which has been studied extensively theoretically~\cite{lim2005, tillmann2013, khalid2017}. This simple illustration indicates that our implementation has the potential to realize such complex multipartite interference.  

\begin{figure}[!t]
\centering
\includegraphics[width=\columnwidth]{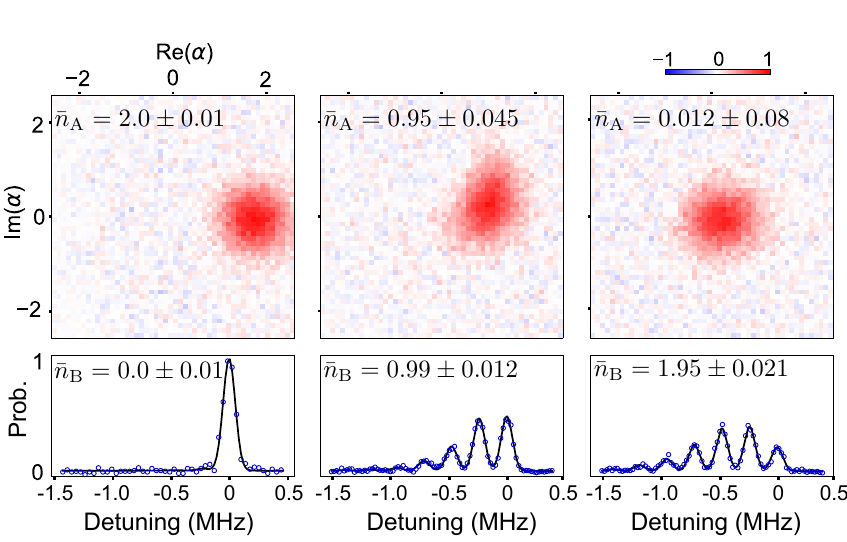}
\caption{\textbf{Behaviour of a coherent state after the BS operation}. The system is prepared in  $|\alpha, 0\rangle_\mathrm{AB}$ with $\alpha=\sqrt{2}$. Top row: measured Wigner function for A under different conditions (a) initial state; (b) 50:50 BS, (c) two 50:50 BS. Bottom row: corresponding photon population measured in B for each conditions respectively}\label{sfig:coherent_state}
\end{figure}

An example of a multiphoton state that can be prepared trivially in the cavities is the coherent state. Here we prepare the system in $|\alpha, 0\rangle_\mathrm{AB}$, with $\alpha = \sqrt{2}$ and measure the population after the operation. We shown in Fig.~\ref{sfig:coherent_state}, the measurement of the initial state, the outcome after a single BS and that after two consecutive BS, which is equivalent to a SWAP operation. Since we only have independent tomography capability on Alice, we chose to measure its Wigner function after each operation using qA and using qC to probe the population in Bob. We can fit both measurements to extract the photon numbers in each cavity. After a single BS, we observe that the coherent state has been reduced to half its original size in Alice. The corresponding measurement of Bob shows that it now contains the same coherent state with $\bar{n}\approx 1$. When we implement the BS operation twice, Alice is fully evacuated to the vacuum state and Bob contains $\bar{n} \approx 2$. This demonstrates that the operation has indeed fully transferred the excitations from Alice to Bob. 

This is a interesting illustration because it highlights the behavior of a semi-classical state under a 50:50 BS. It simply becomes two separable coherent states each of half the size of the initial state. No entanglement is created between Alice and Bob because the BS operation is linear. Further, the presence of larger photon number states also highlights the effects of non-linearities in our system during the operation. We can see this effect from Fig.~\ref{sfig:coherent_state}(b) where the coherent state in Alice after a single BS is distorted due to self-Kerr~\cite{kirchmair2013}. Such non-linear effects can cause complications as we move towards interference between more excitations. Therefore, it is desirable to realize an effective coupling rate much faster than the self-Kerr. More over, the coherent state can simply be considered as a weighted superposition of Fock states. Thus, another implication of this experiment is that our engineered operation is fully capable of handling not only multiple excitations but more importantly, the superpositions of different photon number states. This property is highly desirable since continuous-variable based quantum error correction schemes, such as the cat code~\cite{mirrahimi2014}, are a promising route towards realizing error-protected logical qubits~\cite{ofek2016}. In fact, the engineered bilinear coupling between high-Q modes is an important ingredient in constructing logical gates between multiple bosonic logical qubits encoded superconducting cavities~\cite{mirrahimi2014}. 

\section{Non-idealities at large photon numbers }
So far, we have shown that techniques described in this work are compatible with multi-photon states. However, there are additional sources of imperfections when higher photon number states are present. Additionally, we have utilized the engineered BS and the parity mapping protocol described in Ref.~\cite{sun2014} to probe the quantum state overlap between Alice and Bob (Fig.~4(b)). Experimentally, we observe a reduced contrast of the measured $\langle \hat{P_A}\rangle$ as the photon number population increases. This arises from both the errors in the parity measurement as well as imperfections in the BS operation at higher photon numbers. 

\begin{figure}[b]
\centering
\includegraphics[scale=1]{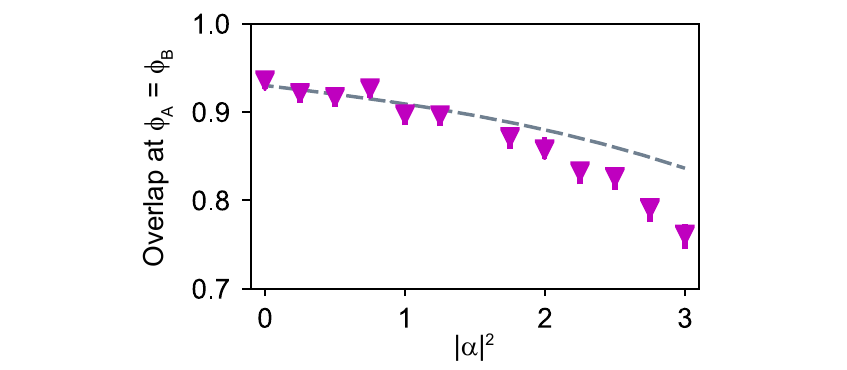}
\caption{\textbf{Measured maximum overlap at different displacements}. We measure the maximum quantum state overlap (magenta triangles) between $|\psi\rangle_{A} = e^{i\phi}|\alpha\rangle$ and $|\psi\rangle_{B} = e^{i\phi_{B}}|\alpha\rangle$ at $\phi_{A} = \phi_{B}$ as a function of $|\alpha|^2$. We observe a reduction in the contrast at larger $\alpha$ due to both the non-linearities of the memories as well as the imperfections of the parity measurement. This is in agreement with the simulation (dashed grey line) where we take into account the self-Kerr, $T_1$, and $T_2$ of each memory mode. }\label{sfig:overlap_contrast}
\end{figure}

A major source of error in the parity measurement at higher photon numbers is the bandwidth of the transmon pulses used in the parity mapping sequence. Ideally, the transmon rotations should have infinite bandwidth such that they are completely independent of cavity photon number. However, this would result in spectral overlaps with other transmon transitions which would cause leakage errors. In light of these two conflicting requirements, we chose to use a Gaussian pulse with $\sigma = 2\pi\cdot 20$\,MHz for the transmon rotations with standard first order DRAG corrections~\cite{motzoi2009} as a compromise. The measured overlap between $|\psi\rangle_{A} = e^{i\phi_A}|\alpha\rangle$ and $|\psi\rangle_{B} = e^{i\phi_{B}}|\alpha\rangle$ at $\phi_{A} = \phi_{B}$ as a function of the $|\alpha|^2$ is shown in Fig.~\ref{sfig:overlap_contrast}. The maximum contrast at $\alpha = 0$ is $\sim$\,94\%, limited by $\sim$\,2\% readout errors and $\sim$\,4\% parity mapping errors due to transmon decoherence. 

The effects of higher photon number states on the BS operation are three-fold.  Firstly, because of the non-linearities of cavities, the frequencies of transitions between excited states of the cavities are detuned from the frequency matching condition for the BS operation, therefore causing reduction in the BS fidelity. Such detunings are proportional to $N\chi_{aa}$, $N\chi_{bb}$, and $N\chi_{ab}$, with $\chi_{aa}$, $\chi_{bb}$ being the self-Kerr of Alice and Bob, and $\chi_{ab}$ the cross-Kerr. We estimate an upper bound on the resulting infidelity to SWAP a Fock state $|N\rangle$ to be proportional to $N[N(\chi_{aa} + \chi_{ab} + \chi_{ab})/g]^2$. Although the intrinsic self-Kerr is relatively small for both Alice and Bob, it is enhanced when the drives are present. This is consistent with distortion observed in Fig.~\ref{sfig:coherent_state}(b). We extract the self-Kerr based on the Wigner tomography of a coherent state after a single BS to be $\chi_{aa}/2\pi \approx 8$\,kHz and $\chi_{bb}/2\pi \approx 5$\,kHz. This drive-induced non-linearity is very sensitive to the exact power and frequency configurations of the drives. A more thorough theoretical and experimental analysis of the change of self-Kerr as a function of drive powers and frequencies is currently underway. With these values of self-Kerr and the independently measured cavity decoherence times, we simulate the maximum overlap measured at $\phi_A = \phi_B$ and find that it is consistent with the measured reduction up to $|\alpha|^2\sim 2$. The additional reduction could be due to the limited spectral bandwidth of the parity measured described in the previous paragraph and other decay mechanisms discussed below.  

Secondly, the decay rates for transitions between the $N$-th and $(N-1)$-th level of the cavities grow as $N$ increases. If the cavity only suffers from linear decay (one photon loss) which is the dominant decay mechanism, the rate of transitions between higher levels increases linearly in $N$. The resulting BS infidelity will scale as $N\kappa/g$ where $\kappa$ is the linear decay rate of the cavities. However, there can be additional loss mechanisms that do not scale as $N$ due to the coupling to a non-linear transmon mode. This leads to the nonlinear decay of the memory modes, i.e., the total rate of decay depends on the instantaneous energy of the cavities: $\kappa_\mathrm{total} = \kappa + N\kappa_{\mathrm{NL}}$. One example of nonlinear decay is two-photon loss where cavities decay by emitting two photons at a time. In general, as the transmon-cavity detuning is large compared to their coupling strength, such non-linear decay is typically weaker than the transmon-induced linear decay and the self-Kerr of the cavities. However, in the presence of drives, the cavity frequencies can be close to certain two-photon transitions where the non-linear decay rates will be enhanced. In this case, the decay rate of transitions between the $N$-th and $(N-1)$-th level of cavities grows as $N^2$ rather than $N$, resulting in an increase in the BS infidelity proportional to $N^2\kappa_{\mathrm{NL}}/g$. 

Thirdly, the transmon is more likely to be excited at large cavity photon number. When the frequency of one of the cavities is close to the resonance frequencies $\nu_{KK'}$ in Eq.~\ref{eq:kappa_gamma}, the probability of exciting the transmon by absorbing both cavity and pump photons increases. The excitation rate is proportional to the cavity photon number $N$. This results in additional non-idealities in the operation when large photon number states are present. In practice, all three of the above-mentioned effects, and potentially other imperfections not considered here, contribute to the degradation in the BS quality. Additional experimental investigations are currently underway in order to provide a more thorough and quantitative analysis. 

\section{Experiment wiring}
\begin{figure*}
\centering
\includegraphics[scale=1]{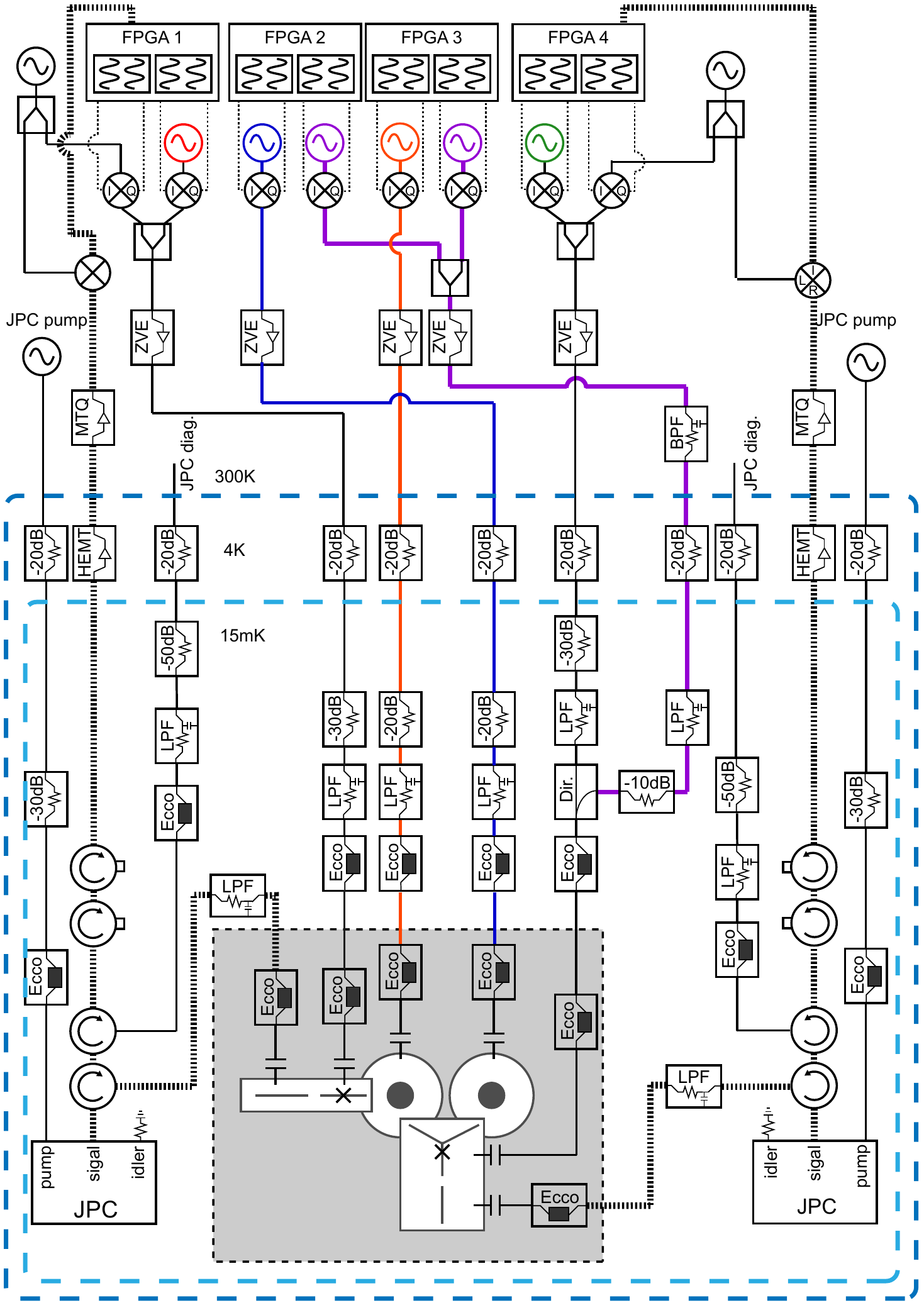}
\caption{\textbf{Room temperature RF controls and fridge line configurations}}
\label{sfig:wiring}
\end{figure*}

The details of the room temperature control configuration and the fridge wirings are shown in Fig~\ref{sfig:wiring}. The device is housed inside a Cryoperm magnetic shield and thermalized to the base plate of a dilution refrigerator with a base temperature of $\sim$15\,mK. We use commercial low-pass (LPF) and custom infrared (Ecco) filters along the microwave lines to reduce stray radiation and photon shot noise. Two Josephson parametric converters (JPCs) are also installed on the base plate. They are connected to output ports of the two readout resonators via circulators and provide near-quantum-limited amplification of the output signal, which is further amplified by commercial HEMT amplifiers at the 4K stage. This gives us the capability to perform efficient single-shot readout of both transmon ancillae independently. 

RF signals used to control each mode are IQ-modulated by DAC outputs from an integrated FPGA system at room temperature. The signal is then mixed with their respective local oscillators (LOs) and then amplified by standard room temperature amplifiers (ZVE and MTQ). Fast switches are placed on the input lines before the signal is transmitted to the fridge in order suppress the unwanted RF power during idle times. The same LO generator is used for the displacement and drive tones for each cavity mode to eliminate relative drifts of their phases. The two drives used to enable the bilinear coupling are combined after their respective amplification and filtered carefully at room temperature. Bandpass filters (BPF) are employed on this line to reduce the spectral width of the noise sent into the fridge. In particular, we suppress the noise near qC resonance frequency by $\gtrsim$ 50 dB to prevent spurious population of the mode. Fast switches are use on all room temperature input lines before they go in to the fridge to prevent excess power input during idle times. 

A dedicated input line is used to bring the drives to the sample at base so that a special attenuation configuration can be used. In this case, the drive line (purple) contains a custom 10\,dB reflective attenuator~\cite{andrew2018} at base. It reflects a large fraction of the input power, which gets dissipated at the upper stages of the fridge where more cooling power is available. This allows us to send large amount RF power without heating up the base plate significantly. The drive tones are then combined with the qC input line via a directional coupler, which has a 6 dB insertion loss. With this configuration, we are able to introduce a relative strong coupling and run the experiments described in the main text at a reasonable duty cycle without causing any measurable heating to the system. \\ 


\bibliography{../ref_lib.bib}        